\documentclass[a4paper]{jpconf}
\usepackage{graphicx}
\begin{document}
\title{Recoiling Ion-Channeling in Direct DM Detectors\footnote{Talk presented at the DSU 2011 Conference, KITPC, Beijing, China, Sept 26-30, 2011.}}

\author{Graciela B. Gelmini}

\address{Department of Physics and Astronomy, University of California Los Angeles (UCLA),\\ 475 Portola Plaza, Los
  Angeles, CA 90095, USA}

\ead{gelmini@physics.ucla.edu}

\begin{abstract}
The channeling of the recoiling nucleus in crystalline detectors after a WIMP collision would produce a larger scintillation or ionization signal in direct detection experiments than otherwise expected.  I present estimates of channeling fractions
obtained using analytic models developed from the 1960's onwards to describe  channeling and blocking effects. We find the fractions to be too small to affect the fits to potential WIMP candidates. I also examine the possibility of detecting a daily modulation of the dark matter signal due to channeling.
\end{abstract}

\vspace{-0.6cm}

The  potential importance of the channeling effect for direct  dark matter detection was first pointed out  for NaI(Tl) by Drobyshevski~\cite{Drobyshevski:2007zj}  in 2007 and soon after by the DAMA collaboration~\cite{DAMA}. Usually only a small fraction $E=Q E_R$  of the recoil energy $E_R$ goes to either scintillation or ionization (e.g. $Q_{\rm I}=0.09$ and  $Q_{\rm Na}=0.30$). Channeled recoiling ions instead give all their energy to electrons, so $Q=1$ and they produce a larger ionization or scintillation signal. The
DAMA paper~\cite{DAMA}  gave an estimate of the channeling fractions of recoiling Na and I ions as function of the energy in which the fractions  grew with decreasing energy to be $\simeq 1$ close to 1 keV (see Fig.4.a). With these channeling fractions, the region of light WIMPs (Weakly Interacting Massive Particle with mass 10 GeV or less) compatible with producing an annual modulation as measured by the DAMA collaboration shifted considerably to lower cross sections, by about one order of magnitude (see e.g. \cite{Savage-2009} or \cite{Belli}). Besides channeling being important for fits to WIMP data, it is also important because  it could produce a daily modulation of the signal dependent on the orientation of the crystal with respect to the galaxy, which would thus  be a background free dark matter signature. In fact, as first pointed out in 
Ref.~\cite{Avignone}, Earth's rotation makes the average WIMP velocity change direction with respect to the crystal which produces a daily modulation in the measured recoil energy (equivalent to a modulation of the quenching factors Q), as most recoils successively come in and go out of crystal channels.   
 
 My collaborators  N. Bozorgnia and P. Gondolo and I,  set out more than three years ago  to do an analytic calculation of channeling in dark matter detectors, initially to estimate  amplitudes of daily modulation due to channeling that could be expected,  and realized that the channeling fraction estimates of DAMA did not include ``blocking", an effect which makes channeling fractions of recoiling ions much smaller at low energies~\cite{BGG1, BGG2, BGG3, BGG4}.

Channeling and blocking in crystals refer to the orientation dependence of ion penetration in crystals. ``Channeling" occurs when ions propagating in a crystal along symmetry axes and planes suffer a series of small-angle scatterings  that maintain them in the open ``channels"  in between rows (axial channels) or planes (planar channels) of lattice atoms and thus penetrate much further than non-channeled ions and give all their energy to electrons.  Channeled ions  do not get close to lattice sites, where they would suffer a large-angle scattering which would take them out of the channel. ``Blocking" is the reduction  of the flux of ions originating in lattice sites along symmetry axes and planes due to the shadowing effect of the lattice atoms directly in front of the emitting lattice site (see e.g. the review by D. Gemmell~\cite{Gemmell:1974ub} and references therein).  

 Channeling and blocking effects in crystals are extensively  used in crystallography,  in measurements of short nuclear lifetimes, in the production of polarized beams etc.  However in all these instances the ion energies are much larger  than the keV to 10's of keV expected in WIMPs collisions. There is one instance in which energies are lower: channeling must be avoided in the implantation of   B, P and As atoms  in Si crystals to make circuits~\cite{Hobler}.   
 
 There are two main theoretical tools to model channeling and blocking in crystals. One of them consists of large simulations codes based on binary collision models (MARLOWE, UT-MARLOWE, CRYSTAl...) using Monte-Carlo codes 
 with stochastic methods to locate crystal atoms and determine impact parameters and scattering angles (TRIM-SRIM), some of them commercially available (e.g. from the SILVACO company). The second
 consists of classical analytic models developed in the 1960's and 70's, in particular by Jens Lindhard in  1965~\cite{Lindhard:1965}.  These are the models used by the DAMA collaboration and  us to estimate channeling fractions.
 
 My collaborators and I used in particular Lindhard's model~\cite{Lindhard:1965} supplemented by the planar channel models of Morgan and Van Vliet~\cite{Morgan} and the 1995-1996 work of G. Hobler~\cite{Hobler} on low energy channeling (applied to ion implantation in Si). In these models the  rows and planes of lattice atoms are replaced by continuum  uniformly charged strings and planes,  in which the screened Thomas-Fermi potential  is averaged over a direction
parallel to a row or plane of lattice atoms to find the transverse continuous potential $U$,
either $U_{\rm axial} (r)$ or $U_{\rm planar}(x)$, which depends only on the distance perpendicular to the string or plane, respectively (see Fig.~1.a). 
\begin{figure} 
\includegraphics[width=.47\textwidth]{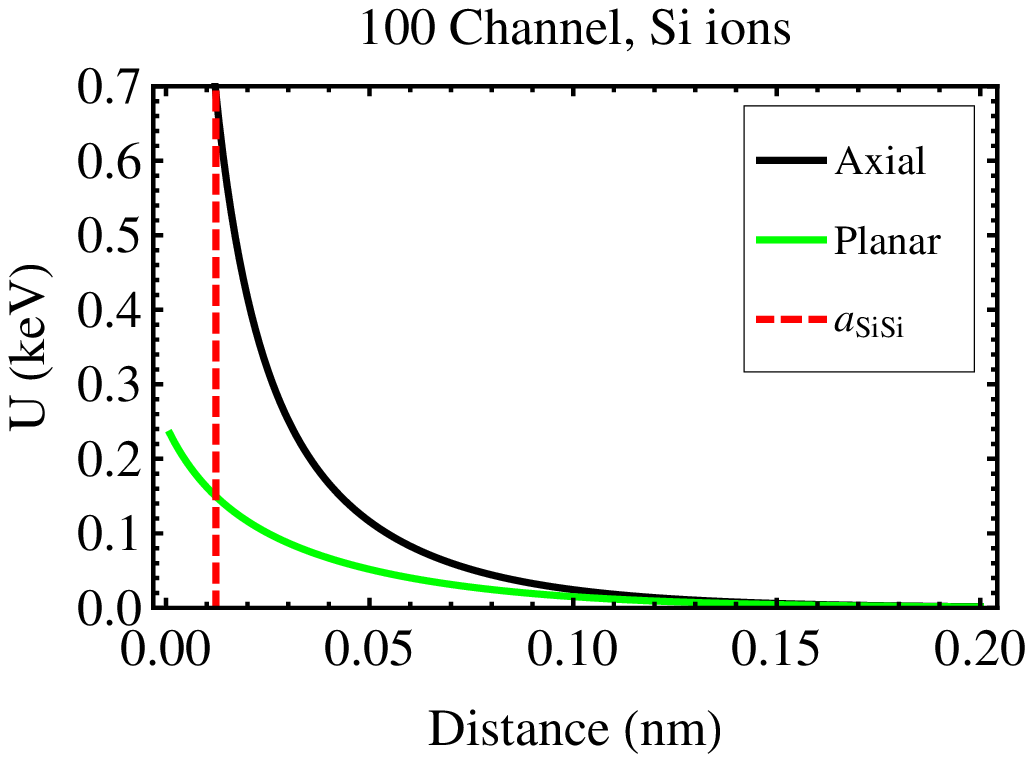} 
\includegraphics[width=.47\textwidth]{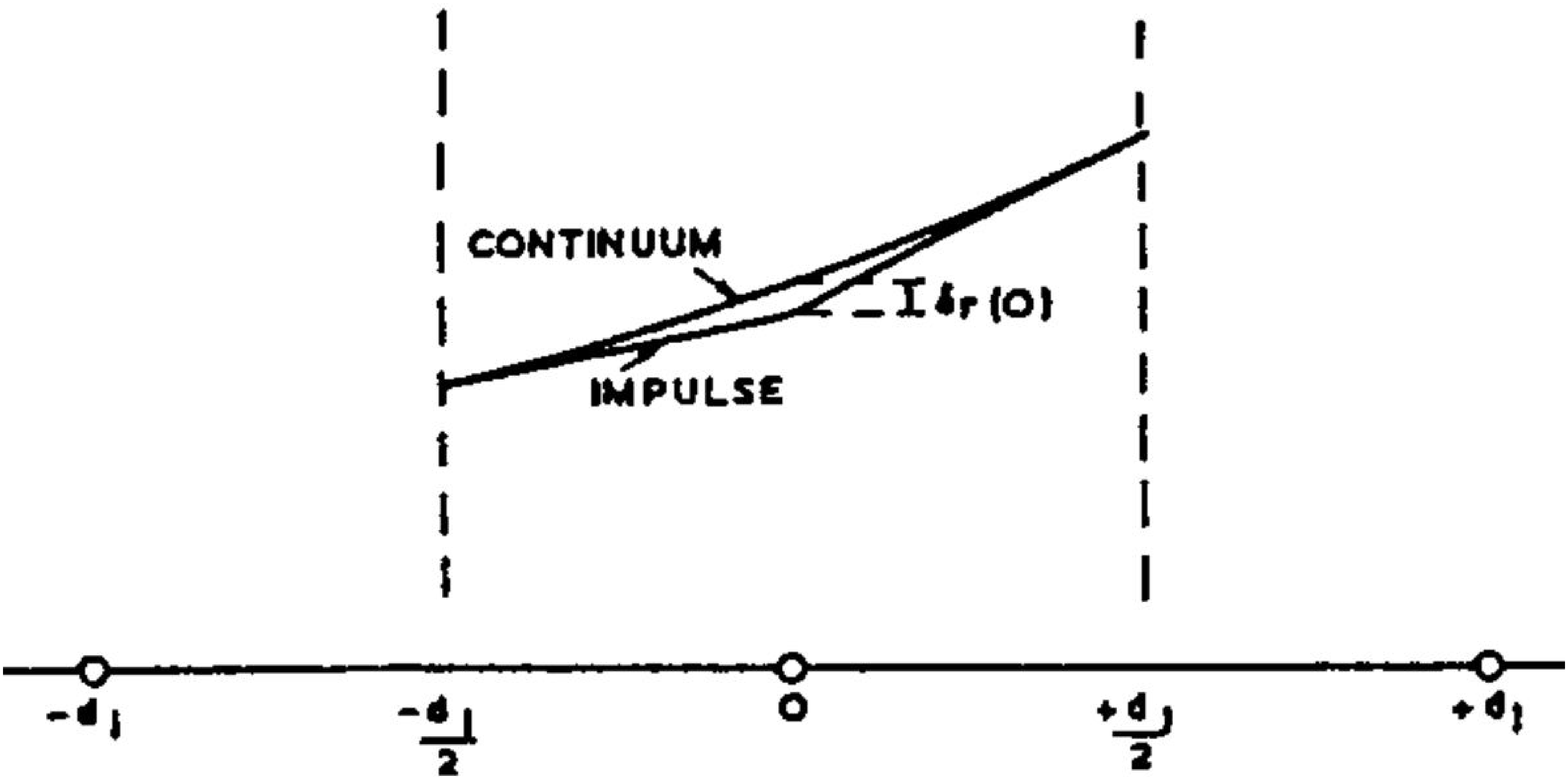} 
\caption{a.(left) Continuum transverse potential $U$ for a string (black) and plane (green) for a Si ion propagating in Si as function of the distance perpendicular to the string or plane~\cite{BGG2}. b.(right) The true ion path lies between those predicted by the continuum and impulse approximations,  schematically shown in the figure taken from Ref.~\cite{Morgan}.  
 The interatomic spacing in the row  is here $d_1$ and two successive half-way planes are shown at $-d_1/2$  and $+d_1/2$.}
\label{fig1} 
\end{figure} 
 The appearance of continuous  strings or planes can be understood as the overlap of  the ``Coulomb shadows" of individual atoms in a lattice row or plane behind the direction of arrival of a parallel beam of positive ions, when the beam arrives at a small enough angle, smaller than a critical angle $\psi_c$. Then, the individual shadows overlap forming a string or plane
  of some thickness within which the scattered ions do not penetrate (see e.g. Fig. 4.6 of Ref.~\cite{ion-book}).
 
   The continuum model  does not imply that the potential energy of an ion moving e.g.  near an atomic row is well  approximated by the continuum potential $U$. The actual potential consists of sharp peaks  near the atoms and deep valleys in between.  The continuum model says that the net deflection due to the succession of impulses from the peaks is identical to the  deflection due to a force $-U'$. This is only so if the ion never approaches any individual atom so closely that it suffers a large-angle collision. Channeled ions must not get closer than a critical distance $\rho_c$  ($r_c$ for an axial channel or  $x_c$ for a planar channel) to the string or plane. We included just one string or plane per channel in our calculation, which is a good approximation except  when $\rho_c$ approaches the  half-width of the channel, which happens only at very low energies.
 
 Along each continuum  uniformly charged string and plane there is translation invariance, thus the parallel component of the propagating channeled  ion velocity  is constant. The only change in energy of the propagating ion is due to the change of the perpendicular component of the velocity ${\rm v}_{\rm perp} = {\rm v} \sin\phi \simeq {\rm v} \phi$, due to the the work of the  forces  $-U'$, which are only perpendicular too. Thus, for channeled ions the so called ``transverse energy" $E_{\rm perp}=  M {\rm v}_{\rm perp}^2/2 + U  \simeq E \phi^2 + U$ is conserved ($M, ~E=M {\rm v}^2/2$ are the mass and kinetic energy of the  ion).

 The conservation of the ``transverse energy" implies that for a particular  value of $E_{\rm perp}$ (determined by the recoil energy $E$, initial recoil angle $\phi_i$ and initial potential  and $U_i$),  a channeled ion does  not approach the string or plane closer than a minimum distance $\rho_{\rm min}$ (for which ${\rm v}_{\rm perp} =0$) and that far away from the string or plane, close to the middle of the channel (where the potential is $ U_{\rm middle}$), the ion moves on a trajectory forming an angle $\psi$ with the string or plane, given by $E_{\rm perp} = E \phi_i^2 + U_i= U(\rho_{\rm min})= E \psi^2  +U_{\rm middle}$. Channeling occurs when $\rho_{\rm min} > \rho_c$, so $U(\rho_{\rm min}) <   U(\rho_c)$, which amounts to 
 \begin{equation}
  \psi=  \sqrt{\left[ U(\rho_{\rm min})- U_{\rm middle})\right]/ E} < \psi_c = \sqrt{\left[ U(\rho_c)- U_{\rm middle})\right]/ E}.
 \label{psi-c}
 \end{equation}
 Here the critical channeling angle far away from the string or plane, $\psi_c$, is  defined in terms of $\rho_c$. All the difficulty of this approach resides in calculating  the critical distances $\rho_c$ for each channel. 
 
 The condition for channeling which defines $\rho_c$ was equated  by Lindhard  in 1965 to the condition that $E_{\rm perp}$  be conserved. For  an axial channel,  Lindhard defined ``half-way" planes perpendicular to the row of lattice atoms half-way in between two successive atoms, and requiring that $E_{\rm perp}$ be the same at leading order in two successive half-way planes (see Appendix A of Ref.~\cite{Lindhard:1965}) obtained the axial channeling condition
 \begin{equation}
 d^2 U''_{\rm axial} (r)/ 8E < 1. 
 \label{axial-cond}
 \end{equation}
 Here $d$ is the interatomic distance in the row considered, $r$ is the radial distance between the propagating ion and the row,  and $U''_{\rm axial} (r)$ is the second derivative of the axial continuum potential with respect to $r$. This condition is fulfilled only at distances $r >r_c$, which defines $r_c$. The same condition for axial channels in Eq.~\ref{axial-cond} was derived later by Morgan and Van Vliet~\cite{Morgan} (with 16 instead of 8) by requiring that the ion position at  two successive half-way planes predicted by the continuum  and the  ``impulse"  approximations be the same, to first order (see Fig. 1.b).    In the impulse approximation the ion suffers an abrupt impulse that changes its momentum only at the closest distance from each lattice atoms. The true situation is in between the impulse and the continuum approximations.

\begin{figure} 
\includegraphics[width=.50\textwidth]{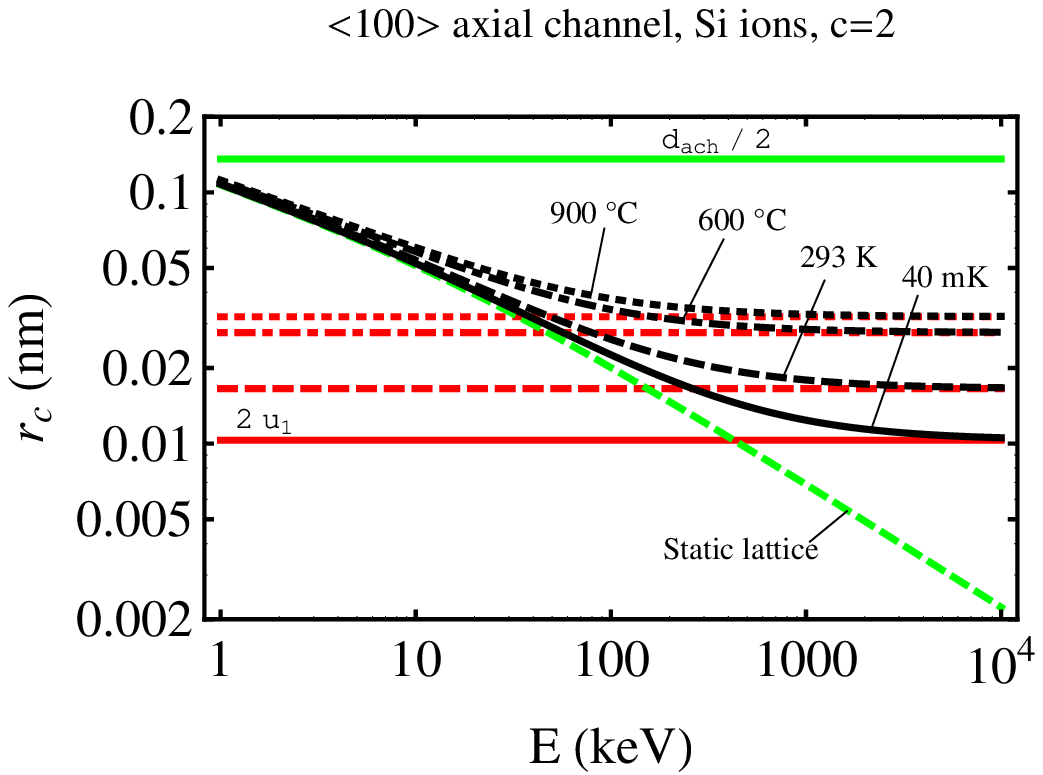} 
\includegraphics[width=.47\textwidth]{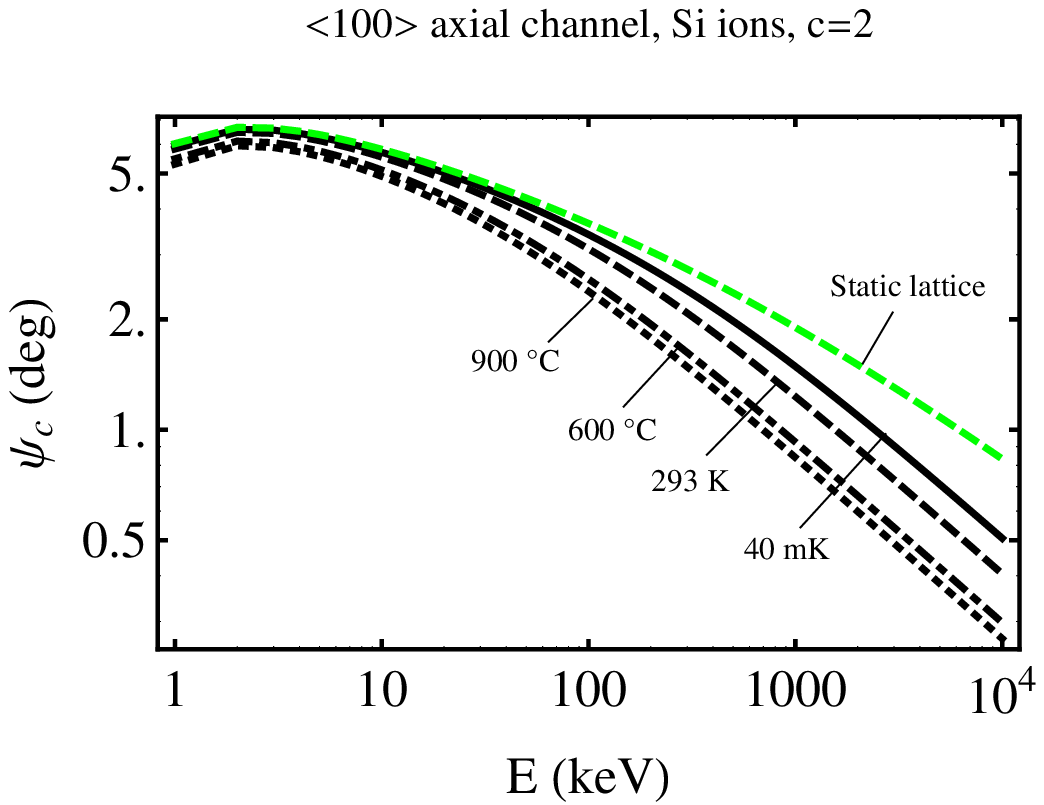} 
\caption{a.(left) Critical channeling distance $r_c$ (Eq.~\ref{critical-dist} where $\rho_c =r_c$ for an axial channel) and b.(right) critical channeling angles $\psi_c$  (Eq.~\ref{psi-c}) for a Si ion propagating in the $<$100$>$ axial channel of a Si crystal, for a static lattice (dashed green) and $T$ corrected (black lines) at different $T$ with $c=2$ (2$u_1$  shown in red and the half width  of the  axial channel, $d_{\rm ach}/2$, in solid green)~\cite{BGG2}.} 
\label{fig2} 
\end{figure} 
\begin{figure} 
\includegraphics[width=.49\textwidth]{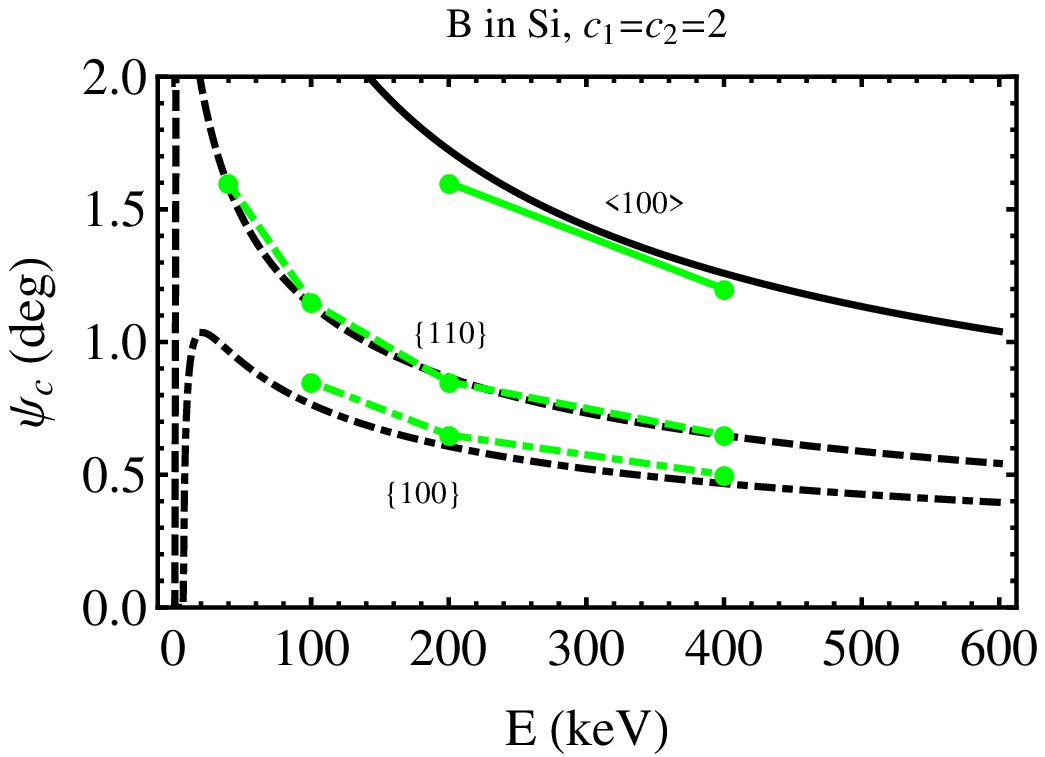} 
\includegraphics[width=.49\textwidth]{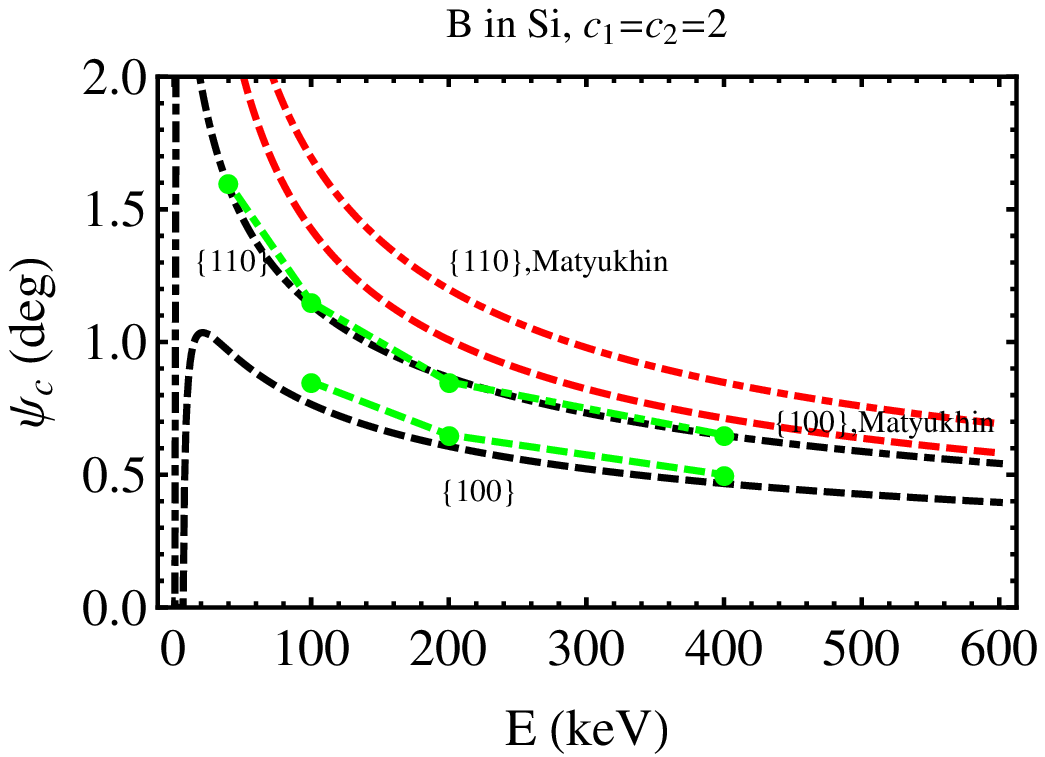} 
\caption{Examples of critical channeling angles $\psi_c$ for B ions propagating in Si  at room T. a.(left)  Our predicted channeling angles using  $c=2$ (black),  are shown here to be very similar to the observed angles given in Ref.~\cite{Hobler} (green) for the  $<$100$>$ axial and the  $\{100\}$ and $\{110\}$  planar channels. b.(right) The corresponding planar critical angles of Ref.~\cite{Matyukhin} (which we believe incorrect)  shown in red, are instead larger.}
\label{fig3} 
\end{figure}

 For planar channels, no condition similar  to Eq.~\ref{axial-cond}  using the second derivative of  $U_{\rm planar} (x)$ can be derived (thus we believe the planar channeling condition of Ref.~\cite{Matyukhin} not to be correct). Planar channels are more complicated than axial channels. To start with, which lattice atoms on a plane are closest  to the projection of the  ion path on the plane depends on the direction of the path itself, and these lattice atoms are in general not at regular distances. For planar channels we followed the approach proposed by Morgan and Van Vliet~\cite{Morgan}, used also by Hobler~\cite{Hobler}, of defining a ``fictitious string" along the projection of the propagating ion path on the plane. 
 
 In the following we call $\rho_c(E)$  the critical distance  for  a static lattice, i.e. one in which vibrations are neglected, obtained from Eq.~\ref{axial-cond} or the equivalent condition using the ``fictitious string" approach for planar channels.  As shown in Fig. 2.a, the static critical distance is a decreasing function of  $E$. However, for our purpose  lattice vibrations cannot be neglected.
 
  The channeling of an ion depends not only on the initial angle its trajectory makes with a string or plane, but also on its initial position. Only for ions starting their motion close to the middle of the channel the channeling condition is simply $\psi <\psi_c$.
However this is not the case in dark matter detectors, since the recoiling ions start their motion at or close to their original lattice sites (and leave those sites empty). For recoiling  ions, blocking effects  are important. In fact in a perfect static lattice no recoiling ion would be channeled, because of what Lindhard called the ``rule of reversibility": due to time reversal symmetry if channeled ions never go very close to lattice sites, the reversed paths do not happen either. It is due to lattice vibrations that the collision with a WIMP may happen while  an atom is displaced with respect to the string or plane where it belongs and, if it is initially far enough into a channel, it may be channeled. The lattice vibrations are strongly temperature ($T$) dependent. Thus $T$ effects must be taken into account in the calculation of channeling fractions. 
They can be incorporated through $T$ dependent critical distances~\cite{Morgan, Hobler} (see Fig.2.a)
 \begin{equation}
\rho_c (E,T) = \sqrt{\rho_c^2(E) +  \left[c~u_1(T)\right]^2}, 
 \label{critical-dist}
 \end{equation}
\begin{figure} 
\includegraphics[width=.52\textwidth]{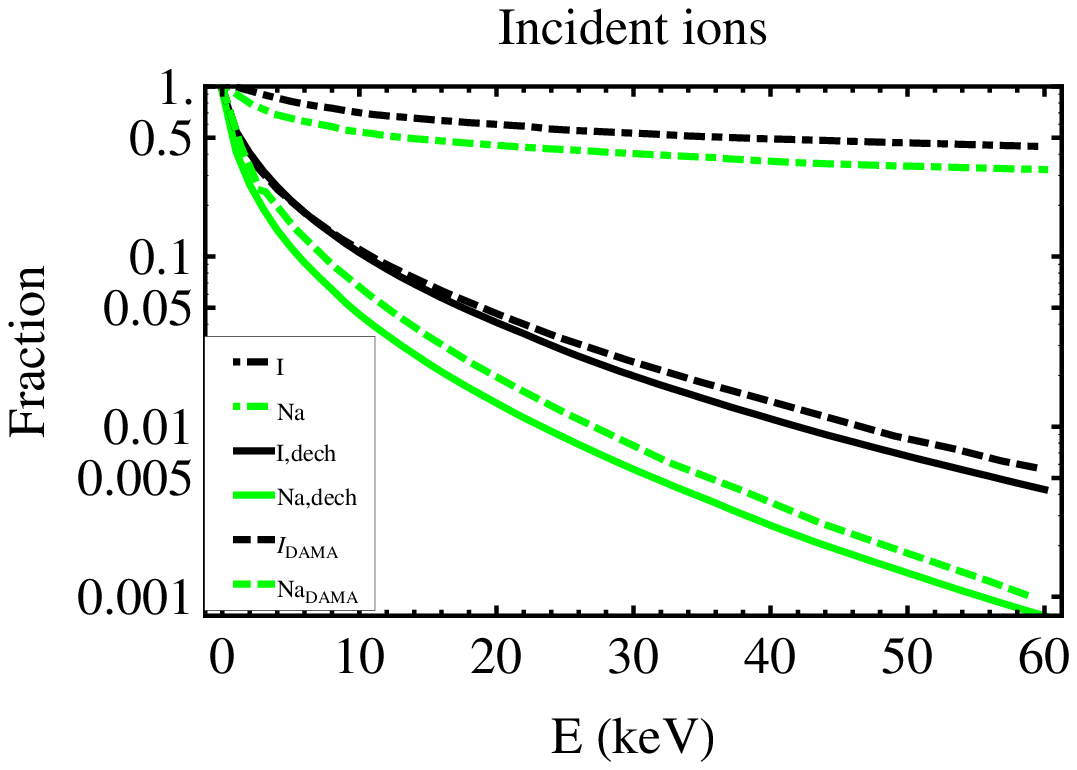} 
\includegraphics[width=.45\textwidth]{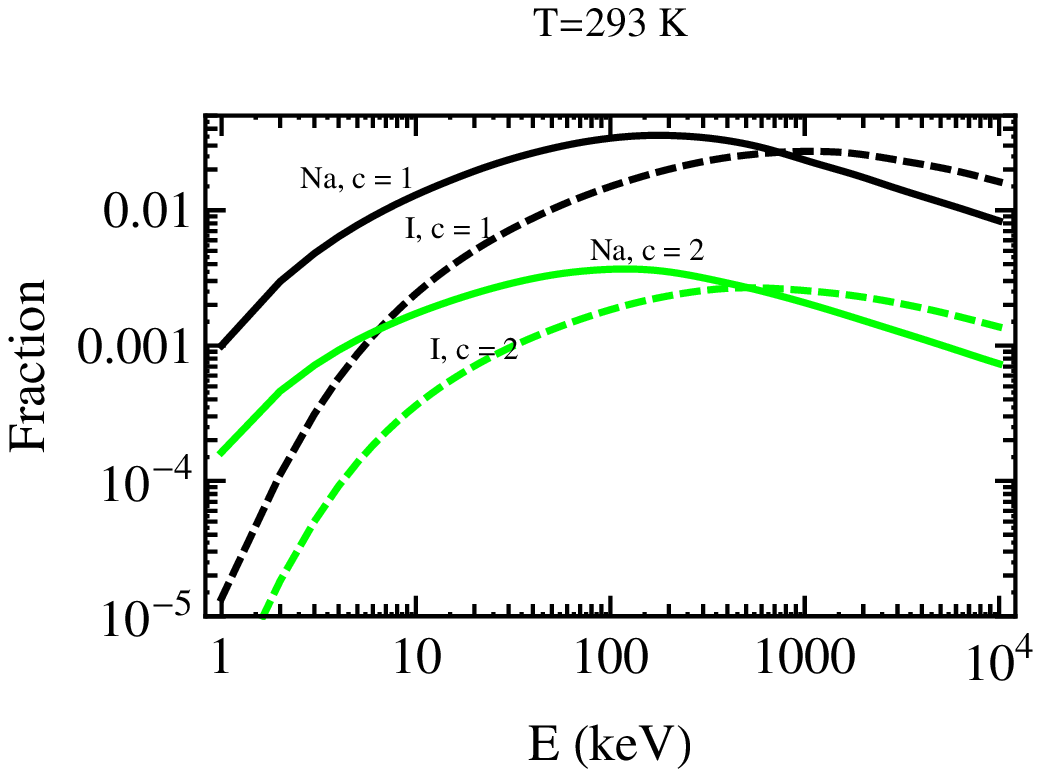} 
\caption{Geometrical channeling fractions of Na and I recoils in NaI(Tl) a.(left) showing the DAMA results~\cite{DAMA} (dashed lines) and ours~\cite{BGG1} (solid lines) neglecting blocking and b.(right) showing upper bounds  to the fractions for the actual situation of ions ejected from lattice sites  at room T and with $c=1$ (green) and $c=2$ (black)
 in Eq. 3~\cite{BGG1}.} 
\label{fig4} 
\end{figure} 
\begin{figure} 
\includegraphics[width=.45\textwidth]{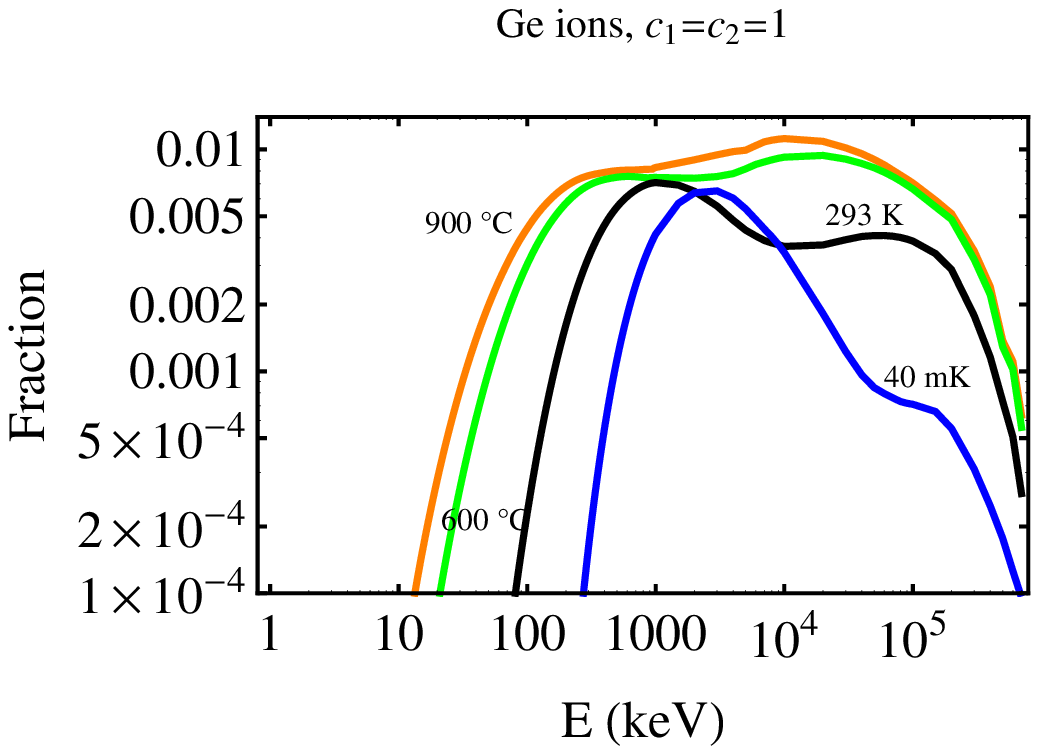} 
\includegraphics[width=.45\textwidth]{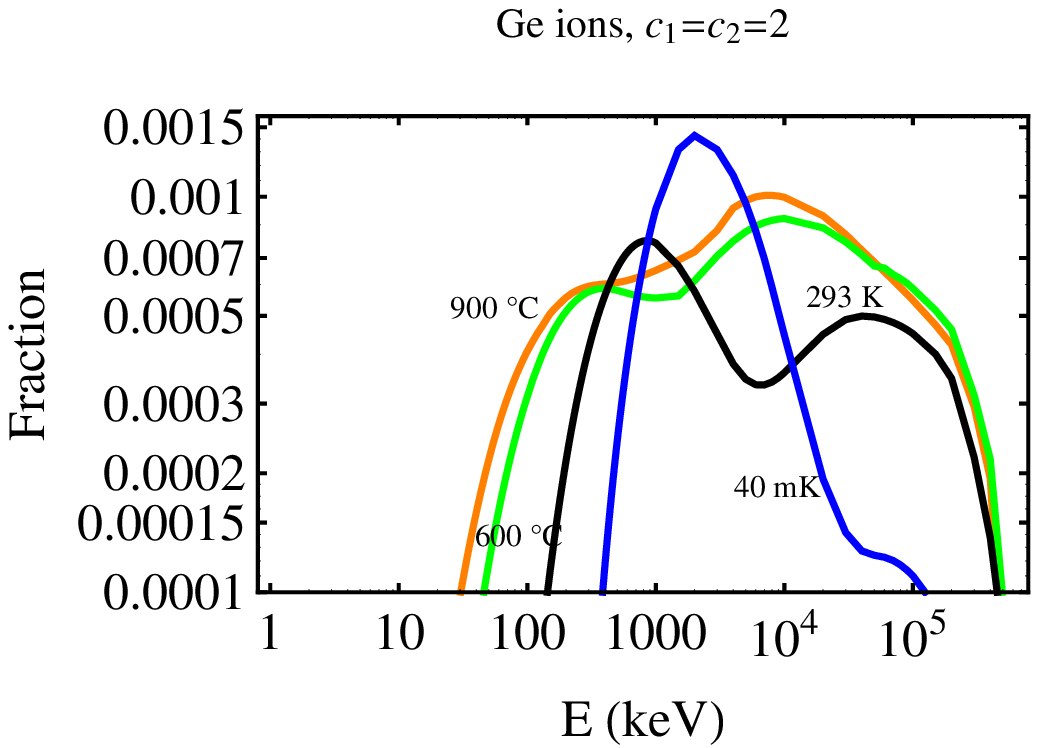} 
\caption{Geometric channeling fractions for Ge ions ejected from lattice sites in a Ge crystal  as function of the ion energy for different  temperatures $T$ and $T$ effects computed (Eq.3) with a.(left) $c=1$ and b.(right) $c=2$.} 
\label{fig 5} 
\end{figure} 
where  $u_1(T)$ is the  1-dimensional amplitude of thermal fluctuations (we used the Debye model) which increases with
 $T$ (see the red lines in Fig.2.a),  and $c$ is a dimensionless number which (for several crystals and propagating ions, different than the ones we study) was found through data and simulations to be $1< c< 2$.  At large enough energies $\rho_c (E,T) \simeq c~u_1(T)$ (see Fig.2.a) and thus  as $T$ increases the strings are planes become thicker, the channels narrower and $\psi_c$ smaller (see Fig.2.b). Using this formalism  with $c=2$ we could reproduce data on channeling angles of B and P ions in Si measured at room $T$ provided by Hobler~\cite{Hobler} (shown in green for B in Figs. 3).  Notice in 
 Fig.3.b that the planar angles of Ref.~\cite{Matyukhin} (derived from a planar channeling condition which, as explained above, we believe is not correct) are larger.

 The critical angles $\psi_c$ increase as the energy decreases, up to an  energy at which they go sharply to zero, signaling that below that particular energy channeling is not possible (see Figs.2.b  and 3). In fact, at those very low energies the  minimum distance from a string or plane required for channeling should be larger than the half-with of the channel. The critical distances  increase as the energy decreases (see  e.g. Fig.2.a) and as they approach the half-width of the respective channel,  the corresponding critical angle $\psi_c$ goes to zero (see Fig. 2.b).

 With our formalism (after including a simple model of dechanneling due to first interactions with Tl impurities) we were able to reproduce  the channeling fractions obtained by DAMA, as shown in Fig.4.a~\cite{BGG1}. 
 These were calculated using only the channeling condition   $\psi < \psi_c$,  i.e. as if the recoiling ions would start their motion far away from the row or plane of atoms.  In this case, we computed the angle that each recoil direction $\hat{\bf q}$ forms with each channel, then assigned a value $\chi(E,\hat{\bf q})=1$  to the probability of channeling in the direction $\hat{\bf q}$ if the angle is smaller than the  critical angle  for any of the channels and zero otherwise. To obtain the total ``geometrical" channeling fraction, i.e. assuming the recoils are isotropically distributed (which in reality they are not - see e.g. Fig.6.b)
 \begin{equation}
P_{\rm geometric}(E)=\frac{1}{4\pi}\int{\chi(E, \hat{\bf q})d\Omega_q}.
 \label{geom-frac}
 \end{equation}
we averaged the channeling probability  over initial recoil  directions. We did it numerically by performing a Riemann sum once the sphere of directions had been divided using the Hierarchical Equal Area isoLatitude Pixelization (HEALPix)~\cite{HEALPix:2005} (a novel use for HEALPix). 

Taking into account that in dark matter detectors the recoiling ions start their motion at or close to their original lattice sites, thus blocking is important, the geometric channeling fractions we obtained are much smaller at low energies (see Fig.4.b). Only if due to lattice vibrations the recoiling  ion is initially far enough  from its original string or plane it may be channeled. Thus, in our model the probability  that an ion is channeled into a particular channel
is given by the fraction of nuclei which can be found at a distance larger than a minimum distance $\rho_{i,\rm min}$ (determined by $\rho_c$ and the initial recoil angle) from the string or plane at the moment of collision. For example, for an axial channel
the probability distribution function of the perpendicular distance to the row of the colliding atom due to thermal vibrations can be represented by a Gaussian, $g(\rho)=(r/u_1^2) \exp(-r^2/2u_1^2)$, thus  the probability  that an ion is channeled is
 \begin{equation}
\chi_{\rm axial}(E,\hat{\bf q}) = \int_{r_{i,\rm min}}^{\infty}{dr g(r)}=\exp{(-r_{i,\rm min}^2/2u_1^2)}
 \label{chi-axial}
 \end{equation}

Notice that any uncertainty in $\rho_c$ is exponentially enhanced in the channeling fraction.  Similar equations apply to planar channels and one needs to compute the probability $\chi(E,\hat{\bf q})$  that an ion with initial energy E is channeled in a given  direction $\hat{\bf q}$ using a recursion of the addition rule in probability theory of the channeling probabilities of all channels~\cite{BGG1, BGG2}.
 To obtain the total geometrical channeling fraction in Eq.~\ref{geom-frac}, we  average the channeling probability  over initial recoil  directions numerically  using HEALPix (as explained above).

  The channeling fractions we obtained are strongly $T$ dependent (see Fig.5). The temperature effects on channeling fractions are complicated. As $T$ increases the probability of finding atoms far from their equilibrium lattice sites increases, which increases the channeling fractions, but the critical distances $\rho_c$ become larger ($\simeq c u_1$ at large enough energies) which decreases the channeling fractions.  The fractions are smaller for larger values of $c$ (see Fig.4.b and compare Fig.5.a and Fig.5.b). Also, we did not include any dechanneling effects. i.e. due to impurities or dopants, which would decrease the channeling fractions. Thus we actually computed upper bounds to the channeling fractions. Using the fractions in Fig. 4.b  with $c=1$ corresponding to the temperature of the DAMA/NaI and the DAMA/LIBRA experiments, T= 293 K, the WIMP regions corresponding to the DAMA annual modulation signal differ only beyond the 5$\sigma$ if channeling is included or not included~\cite{Savage:2010tg}.
 
As mentioned above, channeling is a directional effect which depends on the velocity distribution of WIMPs in the dark halo of our Galaxy and could lead to a daily modulation of the signal~\cite{Avignone}.  The recoil momentum distribution resulting from WIMP collisions is anisotropic, although with a larger dispersion than the WIMP momentum distribution (see Fig.5). Thus Earth's daily rotation changes the direction of the recoil distribution with respect to the crystal axes, changing the amount of recoiling ions that are channeled vs non-channeled. 
\begin{figure} 
\includegraphics[width=.45\textwidth]{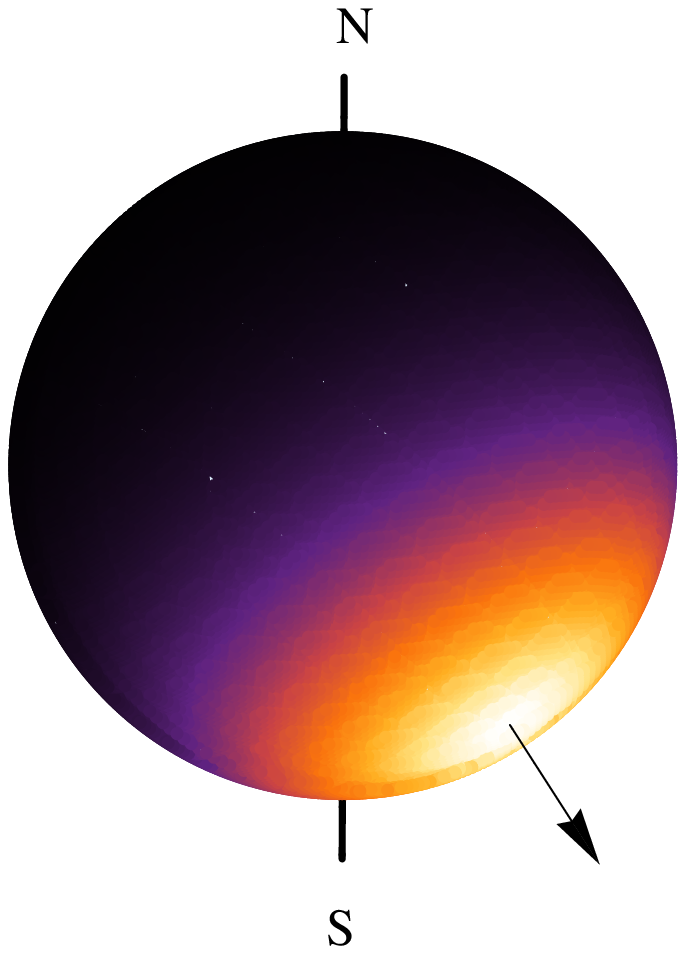} 
\includegraphics[width=.45\textwidth]{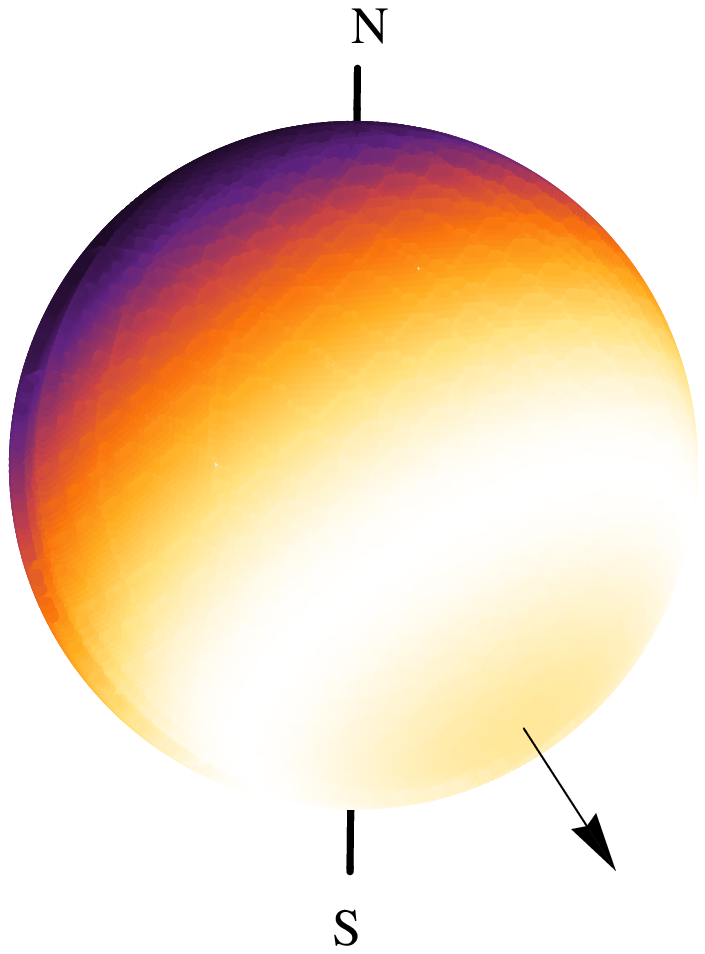} 
\caption{a.(left) Number of 60 GeV WIMPs  per solid angle with speed larger than 196.7 km$/$s (as necessary to produce 10 keV Na recoils)  plotted on  the sphere of WIMP velocity directions and b.(right) number  of 10 keV Na recoils per solid angle plotted on the sphere of recoil directions,  for the standard halo model with average WIMP velocity 288.3 km/s and dispersion 173 km/s. The numbers are normalized to be between 1 (white) and 0 (black). The arrow shows the direction of the average WIMP velocity. Figures from Ref~\cite{BGG-mod}.} 
\label{fig6} 
\end{figure} 
\begin{figure} 
\includegraphics[width=.45\textwidth]{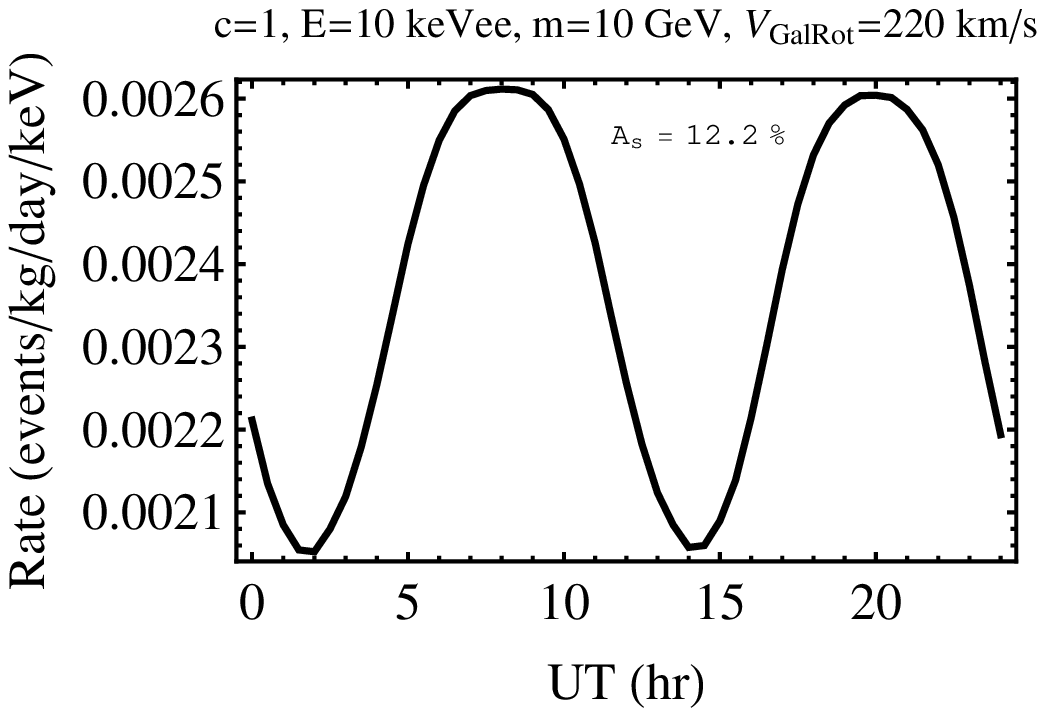} 
\includegraphics[width=.45\textwidth]{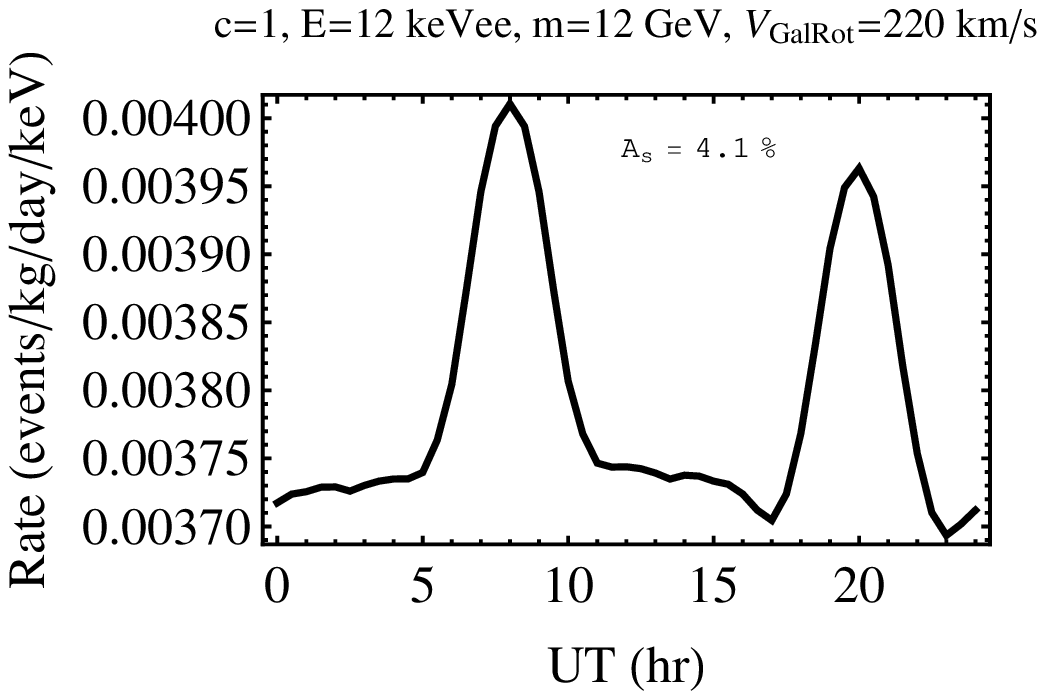} 
\caption{Examples of  the daily variations due to channeling of the signal rate in NaI. Rate (in events/kg-day-keVee) shown as function of the Universal Time (UT) during 24 hours for  WIMP mass and recoil energy  of a.(left)10 GeV and=10 keV, and b.(right) 12 GeV and 12 keV. The parameters used are: 228.4 km/s average WIMP velocity and 300 km/s velocity dispersion, $Q_{\rm Na}=0.2$, $Q_{\rm I}=0.09$, $\sigma_p=2 \times 10^{-40} \textrm{cm}^2$, $c=1$ and $T=293$ K. Figures from Ref~\cite{BGG-mod}.} 
\label{fig7} 
\end{figure} 

In order to compute the daily modulation of the interaction rate we need to orient the crystal with respect to the galaxy and convolute   the  directional recoil rate $dR/dE_R d\Omega_q$, which can be written in terms of the Radon transform~\cite{gondolo-2002} of the WIMP velocity distribution, with the probability $p(E,E_R,\hat{\textbf{q}})dE$ that an energy $E$ is measured when a nucleus recoils in the direction $\hat{\textbf{q}}$ with initial energy $E_R$~\cite{BGG-mod},
 \begin{equation}
\frac{dR}{dE}=\int{{\frac{dR}{dE_R d\Omega_q} {p(E,E_R,\hat{\textbf{q}})}}}d\Omega_q dE_R,
 \label{dRdE}
 \end{equation}
 where  
 \begin{equation}
p(E,E_R,\hat{\textbf{q}})=\chi(E_R, \hat{\textbf{q}})\delta(E-E_R)+[1-\chi(E_R, \hat{\textbf{q}})]\delta(E-QE_R), 
\label{prob}
 \end{equation}
is normalized so that $\int{p(E,E_R,\hat{\textbf{q}})dE}$$=$1. The first term accounts for the channeled events ($E=E_R$) and the second term for the unchanneled events ($E=QE_R$).

After developing the general formalism, in Ref.~\cite{BGG-mod} we examined the possibility of finding a daily modulation due to channeling  in the data already collected by the DAMA/NaI and DAMA/LIBRA experiments. We found that  the signal daily modulation amplitude, defined as $A_s= [R_{s {\rm -max}}-R_{s{\rm -min}}]/[R_{s{\rm-max}}+R_{s{\rm-min}}]$, could be large (of the order of 10\% in some instances)  in NaI crystals (see Fig. 7). However, with a simple observability criterium, we find that
for this daily modulation to be observable  the DAMA total rate should be 1/40 of what it is or the total DAMA exposure should be 40 times larger.  The daily modulation due to channeling will be difficult to measure in future experiments too. As an example, one of the best cases we found~\cite{BGG-mod} is that of a solid Ne detector~\cite{BGG4}, if the WIMP mass is 5 GeV  and  its cross section  is as  large as allowed  by direct detection bounds, $\sigma_p= 10^{-39}$ cm$^2$, then a daily modulation could be detected with a 0.33 ton y exposure, assuming no background.

To conclude, analytic  models of channeling and blocking like those we use give good qualitative results but  channeling data and or simulations  are necessary to get reliable quantitative results  (we used some data available for Si only). Moreover, dechanneling and rechanneling effects cannot be easily computed with these models. Montecarlo simulations may be needed to settle these issues (many are used in other applications of channeling).

\subsection*{Acknowledgments}
G.G.   was supported in part by the US Department of Energy Grant
DE-FG03-91ER40662, Task C. She wants to thank  the APC
(Astroparticles and Cosmology Laboratory), Paris, France, for hospitality while writing this contribution to the  DSU2011 conference proceedings.

\end{document}